# Odd symmetry planar Hall effect: A method of detecting current-induced in-plane magnetization switching

Raghvendra Posti[1], Abhishek Kumar[1], Mayank Baghoria[2], Bhanu Prakash[2], Dhananjay Tiwari[3] and Debangsu Roy[1*]

[1]Department of Physics, Indian Institute of Technology Ropar, Rupnagar 140001, India

[2] Institute of Nano Science and Technology, Phase-10, Sector-64, Mohali, Punjab 160062, India

[3]Advanced safety & User Experience, Aptiv Services, Krakow 30-707, Poland

**ABSTRACT**

Type-x device attracts considerable interest in the field of spintronics due to its robust spin-orbit torque (SOT) induced magnetization switching, and easy deposition technique. However, universally applicable and straightforward detection of type-X magnetization reversal is still elusive, unlike type-Z switching, which employs DC-based anomalous Hall effect measurement. Here, we, demonstrated that the odd planar Hall signal (O-PHV) exhibits an odd symmetry with the application of an external magnetic field which motivates us to develop a reading mechanism for detecting magnetization switching of in-plane magnetized type-X devices. We verified our DC-based reading mechanism in the Pt/Co/NiFe/Pt stack where a thin Co layer is inserted to create dissimilar interfaces about the NiFe layer. Remarkably, the current-induced in-plane fields are found to be significantly large in Pt/Co/NiFe/Pt stack. Further, we successfully employed the O-PHV method to detect the current-induced magnetization switching. The pure DC nature of the writing and reading mechanism of our proposed type-X detection technique through O-PHV makes it the easiest in-plane magnetization detection technique. Moreover, the high repeatability and easy detection of our proposed method will open new avenues toward in-plane SOT switching based memory devices and sensors.

______________________________________________________________________________

[*] Corresponding author: debangsu@iitrpr.ac.in

Search for efficient means to process and store information coupled with negligible dissipation has been a perpetual effort in modern-day electronics. One such goal is to develop magnetoresistive random-access memory (MRAM) operating at high frequency with low energy consumption[1-3]. It was proposed that spin-transfer torque (STT) based MRAM could be an alternative to the presently used field-driven MRAM[4-6]. However, low endurance due to the flow of high current density through the tunnel barrier during the writing operation limits the applicability of STT-MRAM devices[4,7]. In this regard, spin-orbit -torque (SOT) MRAM draws immense attention due to better endurance, faster access time, and lower energy consumption than STT-MRAM[4,7]. In SOT-MRAM, lateral current through the heavy metal (HM) layer generates spin current due to the bulk spin-Hall effect or/and interfacial Rashba-Edelstein effect which induces a torque in the adjacent ferromagnetic (FM) layer leading to the magnetization reversal in the FM layer in HM/FM heterostructures[8-10]. Generally, two orthogonal components, namely, antidamping-like (AD) and field-like (FL) SOT, are realized in the HM/FM heterostructures[11,12].

Depending on the relative orientation of the magnetic easy axis (EA) and spin polarization ($\sigma$) direction, one can envisage three different SOT switching schemes such as type-X (in-plane EA $\perp$ $\sigma$), type-Y (in-plane EA $\parallel$ $\sigma$) and type-Z (out-of-plane EA $\perp$ $\sigma$)[13-17]. It has been shown that for achieving fast switching in type-Y (type–X & type-Z) material, current pulse with a higher (lower) magnitude is required, thus restricting the device's applicability based on type-Y geometry[16,17]. Notably, type-X and type-Z devices show similar magnetization dynamics since, in both cases, EA $\perp$ $\sigma$. Perpendicular magnetic anisotropy [PMA] (in-plane magnetic) based material stacks show type-Z (type- X) switching. Notably, an external magnetic field or equivalent is required to break the symmetry to achieve deterministic switching in both type-Z & type- X devices [13,14,16]. Moreover, deterministic switching in type –X device can be achieved by introducing a slight tilt angle between the current channel and the magnetic easy axis of the in-plane FM layer[16,17]. However, the thin-film deposition and post-processing to achieve an optimized growth of PMA-based heterostructures require rigorous calibration compared to in-plane magnetized stacks. Thus, type-X devices are ideal candidates for durable, faster, and energy-efficient data storage devices and magnetic sensors that can be deposited under easy growth conditions.

Regardless of these benefits, type-X switching is only detectable with limited methods. Presently, two methodologies, namely differential planar Hall Effect technique (we refer this method as 'old PHE')[14,15] and AC 2nd Harmonic technique (we would coin it as 'AC technique')[13] are utilized for detecting type-X switching. In "old PHE", planar Hall signals differing at alternating field values (along DC current direction) define in-plane magnetization direction. For the AC technique, the application of DC is replaced by the sinusoidal AC, and the magnetization state is determined by measuring the 2nd harmonic response of the planar Hall voltage. Both these methods have certain limitations such in 'old PHE', an external alternating magnetic field is required, whereas in 'AC technique', the second harmonic signals are significantly weak, rendering it to be challenging to detect for low-resistance devices. Thus, universally applicable and straightforward detection of type-X magnetization reversal is still lacking, unlike type-Z switching, which employs the measurement of the DC-based anomalous Hall effect (AHE) in presence of an external DC magnetic field[18,19]. In this regard, we propose a more straightforward methodology of detecting type-X switching by measuring a modified PHE signal based on DC measurements in presence of an external DC magnetic field. We have shown our proposed measurement scheme for the in-plane magnetized Pt/Co/NiFe/Pt stack. In this stack, the lateral current passing through it can switch the in-plane magnetization of the stack. We have demonstrated better sensitivity for our proposed scheme than the AC technique. Further, we characterized the SOT-induced effective fields generated by AD-SOT and FL-SOT in our investigated device. Notably, we have engineered this stack by introducing a very thin layer of Co to induce considerable FL-SOT, which would otherwise be absent in stacks where FM has similar HM interfaces[20,21].

AHE ($V_{AHE} \propto m_z$) and PHE ($V_{PHE} \propto m_x \cdot m_y$) signals are the measures of out-of-plane and in-plane magnetization components, respectively. Note that the magnetization switching of the perpendicularly magnetized stack is easy to detect via AHE ($V_{AHE} \propto m_z$) signal. In contrast, the even symmetry of magnetization in PHE ($V_{PHE} \propto m_x \cdot m_y$) makes it difficult to perceive for the in-plane magnetized sample. In this regard, our effort is directed to modify the existing PHE measurement protocol based on the DC and external DC magnetic field so that the resultant quantity based on PHE becomes proportional to the in-plane magnetization.

Generally, for an in-plane magnetized film, the Hall signal has both anomalous and planar Hall contributions, which can be expressed as

$$V_H = V_{AHE} + V_{PHE} = I\Delta R_{AHE} m_z + I\Delta R_{PHE} m_x m_y \qquad (1)$$

Here, $\Delta R_{AHE}$ and $\Delta R_{PHE}$ are AHE and PHE resistances, respectively, and $I$ here is the applied current.

A field sweep along current direction ($H_x$) shows an insignificant contribution from the $m_z$ component for an in-plane magnetized sample. Thus, the AHE contribution became redundant in Eq.1. Thus, Eq. (1) can further be modified as

$$V_H \approx V_{PHE} = I\Delta R_{PHE} \cos\varphi \sin\varphi \qquad (2)$$

Here, $\varphi$ is the azimuthal magnetization angle (illustrated in Fig. 1a). The angle $\varphi$ depends on the resultant magnetic field which generally combines both the external and current-induced field. When a lateral current is passed through the sample (along $\hat{x}$), current-induced AD-SOT and FL-SOT with effective fields of $H_{AD}$ ($\propto \sigma \times m$) and $H_{FL}$ ($\propto \sigma$), respectively (here, $\sigma$ is along $\hat{y}$ and $\boldsymbol{m}$ is along $\hat{x}$) is generated. Here, $H_{AD}$ is along $\hat{z}$ and contributes to the AHE, whereas $H_{FL}$ lies along $\hat{y}$ and modifies the PHE. Nonetheless, current also generates an Oersted field ($H_{Oe}$) along $\hat{y}$, which contributes to the PHE (illustrated in Fig. 1a) as well[22-24]. Thus, the resultant current-induced field takes the following form: $H_I = H_{FL} + H_{Oe}$.

The Taylor series expansion of $\varphi(H_{ext}, H_I)$ about $I = 0$ gives rise to:

$$\varphi(H_{ext}, H_I) = \varphi(H_{ext}, H_I)|_{I=0} + \frac{d\varphi(H_{ext}, H_I)}{dI}(I) + o(h^2)$$

$$\varphi(H_{ext}, H_I) \approx \varphi(H_{ext}, H_I = 0) + \frac{\partial \varphi}{\partial H_I}\frac{\partial H_I}{\partial I}(I) = \varphi_0 + \Delta\varphi$$

$$\Delta\varphi = \frac{\partial \varphi}{\partial H_I}\frac{\partial H_I}{\partial I}I \quad \& \quad \varphi_0 = \varphi(H_{ext}, h_I = 0) \tag{3}$$

Here, $\varphi_0$ is the direction of magnetization before applying the current.

After applying an in-plane current, Eq. (2) can be modified as:

$$V_H \approx V_{PHE} = I\Delta R_{PHE} \cos(\varphi_0 + \Delta\varphi)\sin(\varphi_0 + \Delta\varphi)$$

We have further assumed that the external field required to saturate the magnetization along the in-plane direction is greater than the current-induced field ($\Delta\varphi \ll \varphi_0$). This assumption holds for all the reading currents used in our magnetization detection scheme. Further, the assumption $\Delta\varphi \ll \varphi_0$ implies that $\sin(\Delta\varphi) \approx \Delta\varphi$, and $\cos(\Delta\varphi) \approx 1$. Hence:

$$V_H = I\Delta R_{PHE}\left[\frac{\sin 2\varphi_0}{2} + \Delta\varphi \cos 2\varphi_0 - \frac{\sin 2\varphi_0}{2}(\Delta\varphi)^2\right] \tag{4}$$

Notably, depending on the polarity of the applied DC for the measurement of PHE, the angle $\varphi$ takes the form of $\varphi = \varphi_0 + \Delta\varphi$ for ($+I$) and $\varphi = \varphi_0 - \Delta\varphi$ for ($-I$), respectively. For our magnetization switching detection scheme, we would measure $\Delta V_H = V_H(+I) + V_H(-I)$ which takes the following form by applying Eq. (3) and (4):

$$\Delta V_H = V_H(+I) + V_H(-I) = 2I^2\Delta R_{PHE}\cos 2\varphi_0 \frac{\partial \varphi}{\partial H_I}\frac{\partial H_I}{\partial I} \tag{5}$$

Further, one can approximate $H_I \propto I \Rightarrow \frac{\partial H_I}{\partial I} = k(constant)$) to eq. (5). Moreover, we can estimate $\varphi = \frac{H_I}{H_{ext}}$ for $H_{ext} \parallel \hat{x}$. (With $H_{ext}, I \parallel \hat{x}$, one can expect $H_{FL} + H_{oe}$ is along $\hat{y}$ and $\tan \varphi \approx \varphi = \frac{H_I}{H_{ext}}$)

With the conditions mentioned earlier, Eq. (5) can be reduced to

$$\Delta V_H \propto \frac{1}{H_{ex}} \qquad (6)$$

Similarly,

$$V_H(+I) - V_H(-I) \propto \frac{1}{H_{ex}^2} \qquad (7)$$

Eq. (6) and (7) qualitatively describe that the $\Delta V_H$ have asymmetric nature about the external field, which in turn induces the change in magnetization for the device with in-plane magnetic anisotropy. Afterward, $\Delta V_H$ is coined as odd planar Hall voltage (O-PHV). In contrast, $V_H(+I) - V_H(-I)$ is symmetric. Thus, the estimation of the O-PHV in a sample that exhibits type–X switching would lead to the quantification of the magnetization switching similar to the measurement of AHE for the PMA sample.

To demonstrate the applicability of the O-PHV as a method of detecting the in-plane magnetization switching for the type-X device, a thin film stack of Ta(3)/Pt(3)/Co(0.6)/NiFe(10)/Pt(6) (from hereon we would refer it as NiFe stack) films were deposited onto thermally oxidized Si/SiO$_2$ substrate by dc-magnetron sputtering. The thickness of the films indicated in the parenthesis is nanometers. The deposition was carried out at room temperature with a base vacuum better than 1×10$^{-7}$ Torr and an Ar gas pressure of 3 mTorr. A thin Co layer is added to enhance the FL torque, as discussed later. Subsequently, thin films were patterned into six terminal Hall bar devices with lateral dimensions of 135 × 12 $\mu$m$^2$ using photo-lithography and plasma etching. A device illustration with measurement geometry is depicted in Fig 1a. The current was applied through the current channel along $\hat{x}$, and voltage/resistance was probed along transverse Hall channels. Transport measurements were carried out at room temperature. All DC measurements were carried out using a Keithley 2450 source meter and Keithley 2182A nanovoltmeter. AC harmonic measurements were conducted using a lock-in amplifier (EG&G 7265) at a reference frequency of 577.13 Hz.

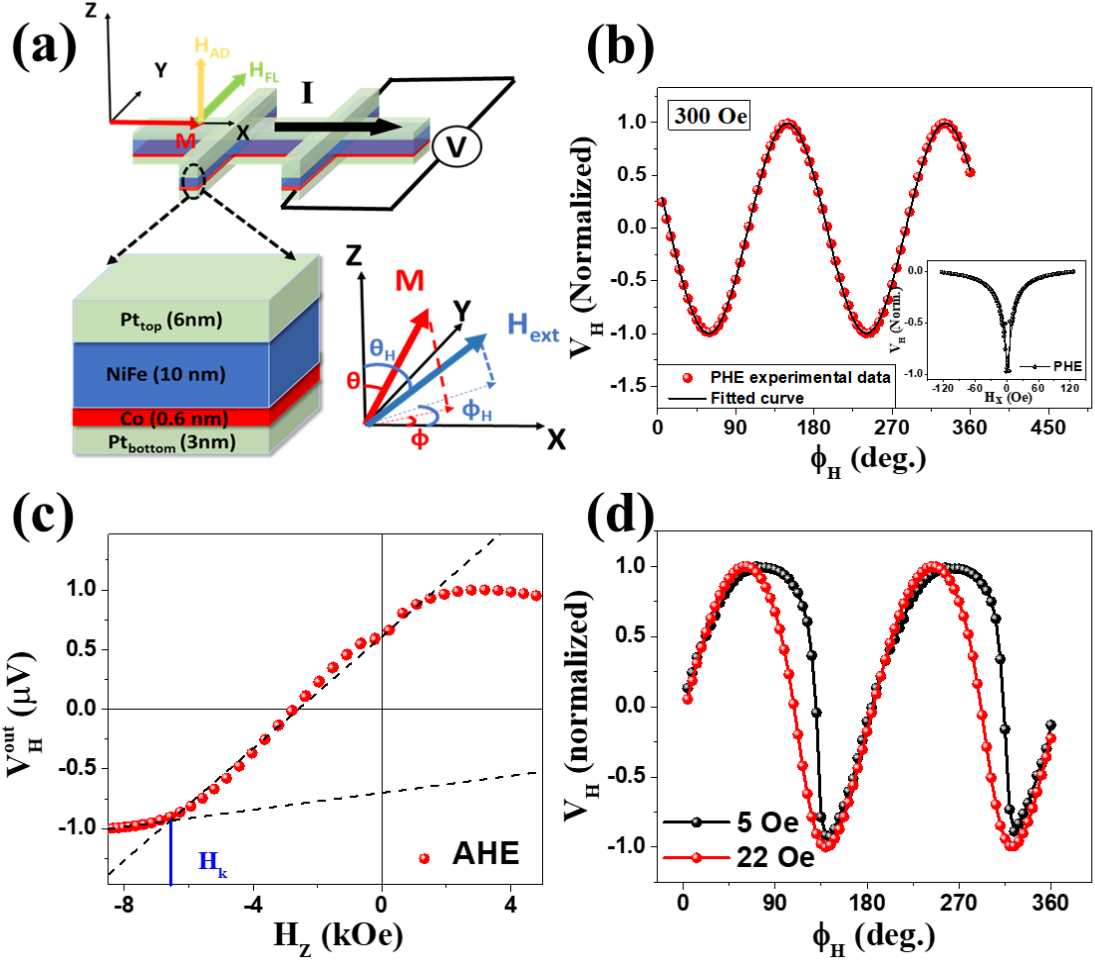

Fig.1 (a) Pt/Co/NiFe/Pt stack with the experimental geometries and SOT directions. (b) Planar Hall voltage ($V_H$) vs. $\phi_H$ in presence of 300 Oe (inset: $V_H$ vs. $H_x$). (c) $V_H^{out}$ vs. $H_z$. (d) $V_H$ vs. $\phi_H$ in the presence of 5 Oe (black) and 22 Oe fields (red).

In-plane magnetic anisotropy present in our stack was confirmed through PHE and AHE experiments (Fig. 1b and Fig 1c). Figure 1(b) depicts planar Hall voltage $(V_H) \propto sin2\phi_H$ dependency for the investigated stack when the sample is rotated in-plane in presence of 300 Oe magnetic field rendering dominating in-plane magnetization component ($V_H \propto m_x \cdot m_y \propto sin2\phi_H$) in our sample. In-plane anisotropy of the sample is further confirmed through the measurement of the variation of $V_H$ as a function of the in-plane field swept along x-direction ($H_x$) (inset of Fig. 1b). The out-of-plane variation of the ($V_H^{out}$) with the varying magnetic field along z-direction is shown in Figure 1(c) which illustrate that the hard axis with anisotropy field ~6587 Oe lies perpendicular to the sample plane. To confirm the in-plane easy direction of magnetization in our stack lies along $\hat{x}$, we have performed the following experiments after saturating the sample magnetization along $\hat{x}$ (by applying 1 kOe external field). The variation of the planar Hall voltage of the NiFe stack was measured by rotating the sample in-plane in presence of two constant DC magnetic fields, 5 Oe & 22 Oe, as shown in Figure 1(d). The angular variation of PHE at ~22 Oe depicts a usual sin2$\phi_H$ dependency, whereas, at 5 Oe, it exhibits a sharp transition at $\phi_H = 90°$ (when *m* is along $\hat{y}$). This confirms the presence of an in-plane magnetic hard axis along $\hat{y}$. Notably, the presence of in-plane easy direction along $\hat{x}$ would further be corroborated in the following section.

Fig. 2a shows $V_H$ for the NiFe stack as a function of $H_x$ by applying ±5 mA DC. Further, O-PHV and $V_H(+5\ mA) - V_H(-5\ mA)$ is estimated from the Figure 2(a), which shows asymmetric and symmetric nature respectively with the field sweep along $H_x$ (Fig. 2b). This validates the qualitative analysis carried out in the preceding section. Furthermore, the hysteric nature of O-PHV in Fig. 2b separates $+m_x$ and $-m_x$ states. To simplify this method, instead of adding the complete PHE curves for *+I* and *-I* (as in Fig 2c), we measured $V_H(+I) + V_H(-I)$ at each value of the magnetic field sweep step (Fig, 2d). We observed a hysteric nature separating two saturation states of $m_x$ making it a useful detection scheme for an in-plane magnetized sample.

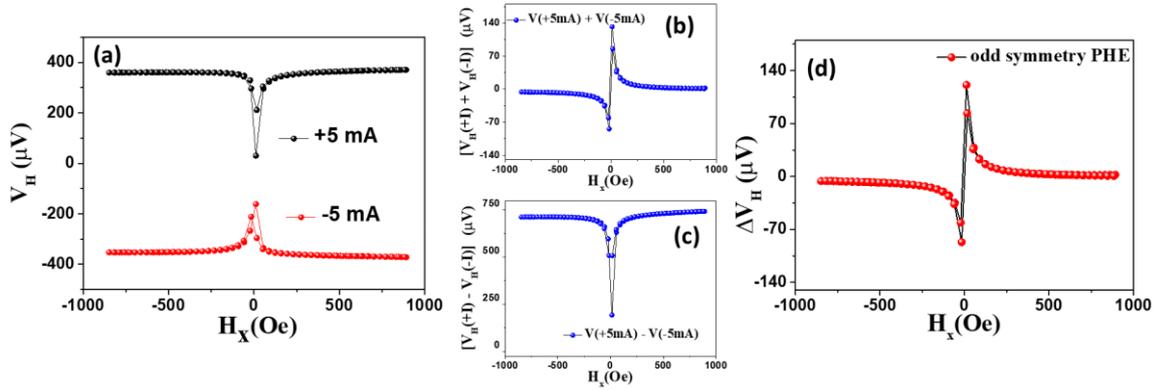

Fig. 2 (a) PHE voltage vs. $H_x$ at ±5 mA currents. (b) $V_H(+5\ mA) - V_H(-5\ mA)$, and (c) $V_H(+5\ mA) + V_H(-5\ mA)$ vs. $H_x$. (d) O-PHV ($\Delta V_H$) measured at magnetic field sweep step.

Based on our calculation and subsequent measurement of O-PHV, one would assume the presence of considerable $H_{FL}$ in the NiFe stack. In order to quantitatively evaluate the contribution of $H_{AD}$ and $H_{FL}$, we have measured the 2$^{nd}$ harmonic contribution of the Hall voltage while sweeping the applied magnetic field along $\hat{x}$ (Figure 3). The experimentally obtained variation of $R_{xy}^{2\omega}\ vs.\ H_x$ was subsequently fitted using the following equation[12,13,22,25].

$$R_{xy}^{2\omega} = \left[\left(-R_{AHE}\frac{H_{AD}}{H_{ext}-H_k} + R_{\nabla T}\right)\cos\varphi + 2R_{PHE}(2\cos^3\varphi - \cos\varphi)\frac{H_{FL}+H_{Oe}}{H_{ext}}\right] \qquad (8)$$

Here, anomalous and planar Hall resistances are $R_{AHE}$ = 109 mΩ and $R_{PHE}$ = 19 mΩ, respectively. The $R_{\nabla T}$ is a signal generated by the anomalous Nernst effect (ANE). ANE is induced by a temperature gradient $\nabla T$, produced by Joule heating. We have repeated this measurement for different magnitude of the AC through the NiFe stack, and the obtained values of $H_{AD}$ and $H_I$ (= $H_{FL}$ + $H_{Oe}$) are plotted for different current amplitudes (Fig. 3a). The rate of change of $H_{AD}$ and $H_I$ with the applied AC is found to be 0.31 Oe/mA and 0.38 Oe/mA, respectively. The ANE signal, as compared to the coefficients of effective field terms, is insignificant[13] in our case ($R_{\nabla T}$~5.9 µΩ/mA). Further, the Oersted field contribution ($H_{Oe}$) is disentangled from $H_I$ in Fig. 3b. The $H_{Oe}$ field originates due to the lateral current flowing through both Pt layers, calculated using[23] $f_{HM}\frac{\mu_0 I}{2w}$. (Here, $f_{HM}$ is the fraction of current ($I$) flowing through the HM layers, $\mu_0$ is vacuum permeability, and $w$ is Hall bar width.). It is found that at a given current value, $H_I$ contribution exceeds $H_{AD}$. Generally, the $H_{AD}$ field is generated through bulk HM whereas the $H_{FL}$ originated due to the inversion symmetry breaking at the interface. It is noteworthy that to achieve considerable $H_{FL}$ in the NiFe stack, which is otherwise absent in an FM layer with symmetric interface[20], we have engineered the NiFe stack by introducing a thin Co layer between Pt and NiFe interface. This leads to the dissimilar interfaces about the NiFe layer resulting in a higher value of $H_{FL}$ (hence, $H_I$).

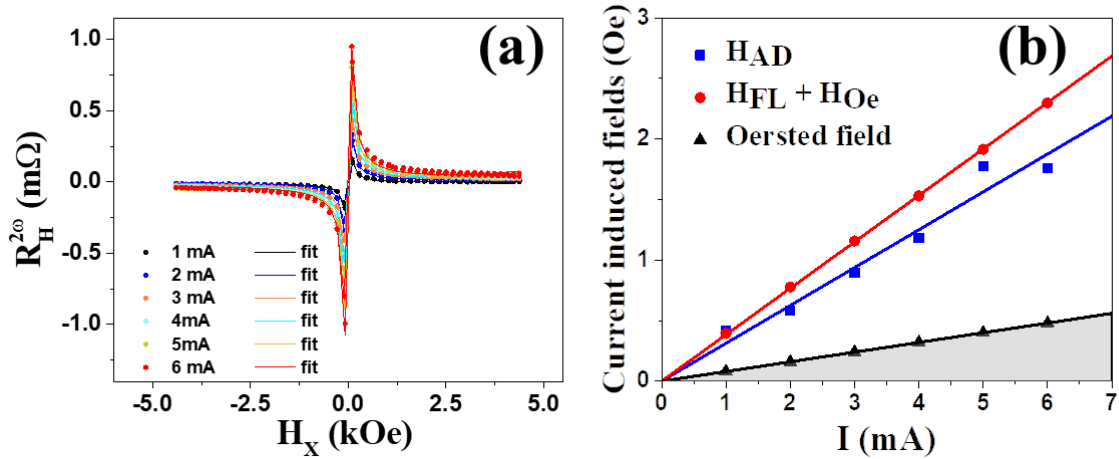

Fig. 3 (a) $R_{xy}^{2\omega}$ vs. $H_x$ for different applied AC (solid circles) current and the fitted curve (corresponding lines) using Eq. 8. (b) $H_{AD}$, $H_I$, and $H_{Oe}$ contributions as a function of applied AC.

Next, we evaluated the NiFe stack's switching behavior in presence of an externally applied field using the 'AC technique' (Fig.4 (a))[13]. In this technique, an alternating current is applied during external magnetic field sweep after each field step, and second harmonic Hall signal ($V_H^{2\omega}$) is detected. The $V_H^{2\omega}$ as a function the in-plane field differentiates the two polarities of in-plane magnetization reversal (Fig. 4 (a)). It is found that the O-PHV based reading mechanism (Fig. 4 (b)) shows an enhanced signal as compared to the AC technique at same current value. Quantitatively, the ratio of DC Hall signal amplitude to the second harmonic Hall signal amplitude shows $[\frac{(\Delta V_H)_{amp}}{(V_{2\omega})_{amp}} = \frac{3.6\ \mu V}{0.6\ \mu V}]$ ~6 times enhanced O-PHV signal at 1 mA read current. Therefore, the $m_x$ states in our proposed reading mechanism are easier to detect than the AC technique. We have estimated the coercivity of the stack using both methods (inset of Fig 4), and it is found to be ~12 Oe consistent for both methods. This further validates our proposed magnetization switching detection scheme using O-PHV.

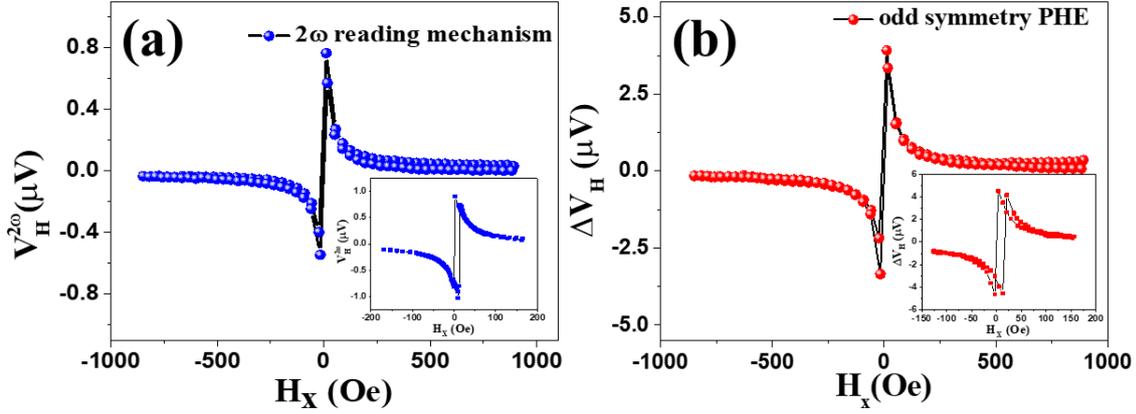

Fig. 4 (a) $V_H^{2\omega}$ vs. $H_x$. (b) O-PHV vs. $H_x$ (at 1 mA applied currents).

Further, we have employed the O-PHV to detect the current-induced magnetization switching in the NiFe. We have applied 1 ms writing current pulses along $\hat{x}$ for the current-induced switching, and the magnetization orientation is subsequently detected by our proposed method (scheme illustrated in Fig 5 (a)). Here the magnetization state was detected using ±1 mA probe current. We have averaged 20 readings to obtain the final data corresponding to a particular writing current pulse for a better signal-to-noise ratio. In presence of a symmetry-breaking field $H_z$, we observed a hysteric behavior (Fig. 5 (b)), which corresponds to current-assisted magnetization switching. We observed similar hysteresis behavior for the negative and positive polarity of $H_z$, corroborating similar results in the literature[13]. Our stack's in-plane switching current was found to be ~24 mA. Without the application of $H_z$, we did not observe the current-induced magnetization switching (Fig. 5 (b)). It is found in previous reports that a small tilt of magnetization from the current channel (in XY-plane) can induce a field-free switching[13,16,17]. However, the absence of switching behavior without applying any symmetry-breaking field suggests no tilt of magnetization from the x-direction. Therefore, as discussed earlier, our devices are truly type-X supporting measurements of angular variation of PHE in Fig. 1d. To check the reproducibility of the current-induced switching using our method, we measured the O-PHV voltage for 200 current cycles of +35 mA and -35 mA pulses. After applying a current pulse (+35 mA), the O-PHV signal was measured in presence of $H_z = +20$ Oe using ±1 mA probe current. The exact process was repeated for the -35 mA current pulse. The occurrence of bipolar resistance states up to 200 current pulse cycles verify the high reproducibility of our proposed method.

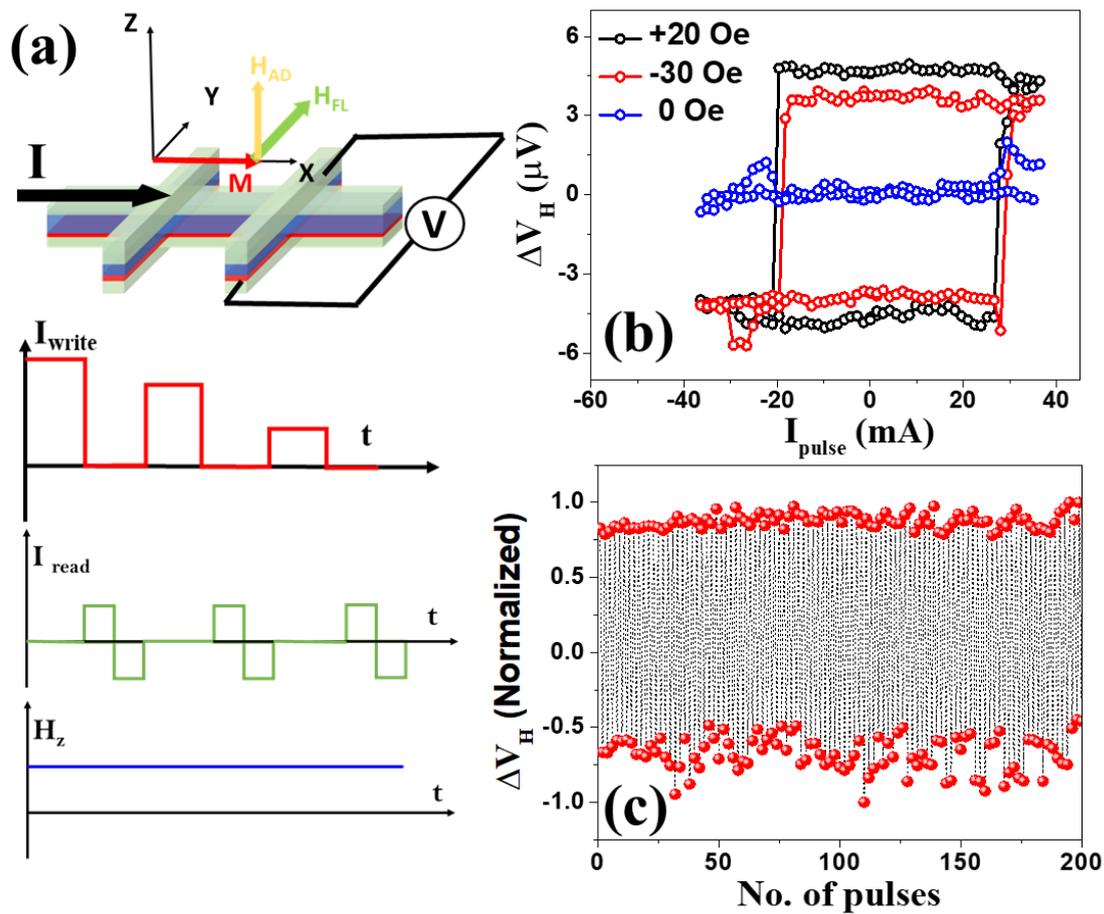

Fig. 5 (a) Measurement scheme for the O-PHV method. (b) O-PHV (read by ±1 mA) as function of DC pulses in the presence of different $H_z$ fields (-30, 0, and, +20 Oe). (c) O-PHV (read currents ±1 mA) after consecutive ±35 mA pulses in presence of 20 Oe $H_z$.

In summary, we showed that the odd planar Hall signal (O-PHV) exhibits an odd symmetry with the application of an external magnetic field. This motivates us to develop a reading mechanism for detecting magnetization switching of in-plane magnetized type-X devices by simply employing the DC technique in line with the widely utilized AHE technique for the type-Z device. We verified our reading mechanism in the Pt/Co/NiFe/Pt stack. We have engineered this stack by inserting a thin layer of Co to create dissimilar interfaces about NiFe layer. This results generation of considerable FL-SOT apart from the expected AD-SOT. Moreover, the resultant of FL-SOT and Oersted field exceeds the AD-SOT's magnitude. It was shown that the O-PHV has higher signal amplitude than the 'AC technique'. Further, we have detected the current-induced magnetization switching in the Pt/Co/NiFe/Pt stack in presence of a symmetry-breaking DC magnetic field. The near-perfect reproducibility of O-PHV in detecting the current-induced magnetization switching in the Pt/Co/NiFe/Pt stack further confirms its applicability in elucidating the in-plane magnetization switching in type-X devices which may lead to its relevance in detecting future SOT-switching-based memory devices and sensors.

DR acknowledges the financial support from the Department of Atomic Energy (DAE) under project no. 58/20/10/2020-BRNS/37125 & Science and Engineering Research Board (SERB) under project no. CRG/2020/005306. AK acknowledges the financial assistance from UGC.

**AUTHOR DECLARATIONS**

**DATA AVAILABILITY**

The data that support the findings of this study are available from corresponding author upon reasonable request.

**References:**

1. N. H. D. Khang, T. Shirokura, T. Fan, M. Takahashi, N. Nakatani, D. Kato, Y. Miyamoto, and P. N. Hai, Applied Physics Letters **120** (15), 152401 (2022).
2. H. Wu, D. Turan, Q. Pan, C.-Y. Yang, G. Wu, S. A. Razavi, B. Dai, N. T. Yardimci, Z. Huang, J. Zhang, Y.-Y. Chin, H.-J. Lin, C.-H. Lai, Z. Zhang, M. Jarrahi, and K. L. Wang, Physical Review Applied **18** (6), 064012 (2022).
3. K. Garello, C. O. Avci, I. M. Miron, M. Baumgartner, A. Ghosh, S. Auffret, O. Boulle, G. Gaudin, and P. Gambardella, Applied Physics Letters **105** (21), 212402 (2014).
4. Y. J. A. b. Huai, **18** (6), 33 (2008).
5. W.-G. Wang, M. Li, S. Hageman, and C. L. Chien, Nature Materials **11** (1), 64 (2012).
6. T. Kawahara, K. Ito, R. Takemura, and H. Ohno, Microelectronics Reliability **52** (4), 613 (2012).
7. A. Manchon, J. Železný, I. M. Miron, T. Jungwirth, J. Sinova, A. Thiaville, K. Garello, and P. Gambardella, Reviews of Modern Physics **91** (3), 035004 (2019).
8. A. Manchon, H. C. Koo, J. Nitta, S. M. Frolov, and R. A. Duine, Nature Materials **14** (9), 871 (2015).
9. J. Sinova, S. O. Valenzuela, J. Wunderlich, C. H. Back, and T. Jungwirth, Reviews of Modern Physics **87** (4), 1213 (2015).
10. J. E. Hirsch, Physical Review Letters **83** (9), 1834 (1999).
11. K. Garello, I. M. Miron, C. O. Avci, F. Freimuth, Y. Mokrousov, S. Blügel, S. Auffret, O. Boulle, G. Gaudin, and P. Gambardella, Nature Nanotechnology **8** (8), 587 (2013).
12. C. O. Avci, K. Garello, M. Gabureac, A. Ghosh, A. Fuhrer, S. F. Alvarado, and P. Gambardella, Physical Review B **90** (22), 224427 (2014).
13. N. H. D. Khang and P. N. Hai, Applied Physics Letters **117** (25), 252402 (2020).
14. Y. Takahashi, Y. Takeuchi, C. Zhang, B. Jinnai, S. Fukami, and H. Ohno, Applied Physics Letters **114** (1), 012410 (2019).
15. G. Mihajlović, O. Mosendz, L. Wan, N. Smith, Y. Choi, Y. Wang, and J. A. Katine, Applied Physics Letters **109** (19), 192404 (2016).
16. S. Fukami, T. Anekawa, C. Zhang, and H. Ohno, Nature Nanotechnology **11** (7), 621 (2016).
17. Y.-T. Liu, C.-C. Huang, K.-H. Chen, Y.-H. Huang, C.-C. Tsai, T.-Y. Chang, and C.-F. Pai, Physical Review Applied **16** (2), 024021 (2021).

18    L. Liu, O. J. Lee, T. J. Gudmundsen, D. C. Ralph, and R. A. Buhrman,  Physical Review Letters **109** (9), 096602 (2012).
19    L. Liu, C.-F. Pai, Y. Li, H. W. Tseng, D. C. Ralph, and R. A. Buhrman,  Science **336** (6081), 555 (2012).
20    R. Posti, A. Kumar, D. Tiwari, and D. Roy,  Applied Physics Letters **121** (22), 223502 (2022).
21    S. Li and T. Zhu,  Japanese Journal of Applied Physics **59** (4), 040906 (2020).
22    F. Xue, S.-J. Lin, P. Li, W. Hwang, Y.-L. Huang, W. Tsai, and S. X. Wang,  APL Materials **9** (10), 101106 (2021).
23    S. Emori, T. Nan, A. M. Belkessam, X. Wang, A. D. Matyushov, C. J. Babroski, Y. Gao, H. Lin, and N. X. Sun,  Physical Review B **93** (18), 180402 (2016).
24    X. Fan, J. Wu, Y. Chen, M. J. Jerry, H. Zhang, and J. Q. Xiao,  Nature Communications **4** (1), 1799 (2013).
25    A. Ghosh, K. Garello, C. O. Avci, M. Gabureac, and P. Gambardella,  Physical Review Applied **7** (1), 014004 (2017).